# Parametric Study of Nonlinear Adaptive Cruise Control for a Road Vehicle Model by MPC

ZEESHAN ALI MEMON*, MUKHTIAR ALI UNAR**, AND DUR MUHAMMAD PATHAN*

## ABSTRACT

MPC (Model Predictive Control) techniques, with constraints, are applied to a nonlinear vehicle model for the development of an ACC (Adaptive Cruise Control) system for transitional manoeuvres. The dynamic model of the vehicle is developed in the continuous-time domain and captures the real dynamics of the sub-vehicle models for steady-state and transient operations. A parametric study for the MPC method is conducted to analyse the response of the ACC vehicle for critical manoeuvres. The simulation results show the significant sensitivity of the response of the vehicle model with ACC to controller parameter and comparisons are made with a previous study. Furthermore, the approach adopted in this work is believed to reflect the control actions taken by a real vehicle.

## 1.    INTRODUCTION

An application of mathematical control techniques to the longitudinal dynamics of a road vehicle with an ACC system has been presented to address vehicle control. ACC systems have been developed as an enhancement to the standard cruise control systems. The ACC system operates on the throttle as well as brakes to maintain a desired speed and a SIVD (Specified Inter-Vehicle Distance) from a preceding vehicle in its vehicle-following mode. An ACC system typically aims to increase road safety and passenger comfort.

A number of ACC vehicle models and controller approaches have been developed in the literature which cover a wide range of ACC vehicle applications. The vehicle models used range from the simple vehicle model, which does not take into account the engine and drive-train dynamics, to the nonlinear vehicle models. The simple vehicle models used are the longitudinal vehicle model [1-6], and first-order vehicle models [7,8]. In either case, the input to the simple ACC vehicle model is the control signal calculated by the ULC (Upper Level Controller). Simple ACC-vehicle models have been used in the previous studies to analyse the performance of the ULC. In the case of a nonlinear vehicle model, the desired acceleration commands obtained from the ULC are given to the LLC (Lower-Level Controller) which then computes the required throttle and brake commands for the nonlinear vehicle model to follow the required acceleration commands. The nonlinear model includes the vehicle engine model, transmission model, wheel model, brake model, ULC and LLC models. In the literature, various control algorithms

*        Assistant Professor, Department of Mechanical Engineering, Mehran University of Engineering & Technology, Jamshoro.
**      Professor, Department of Comuter Systems Engineering, Mehran University of Engineering & Technology, Jamshoro.

have been developed for the ULC, namely, PID (Proportional Integral Derivative) control [9,10], sliding mode control [5,6,11-14], CTG (Constant Time Gap) [7,8], and MPC [7,15-18].

## 1.1 Transitional Manoeuvres for Accident Avoidance

It is not always necessary that an ACC vehicle has to perform steady-state operations [1,5,17,19]. It might need to execute TMs (Transitional Manoeuvres), e.g. it might encounter a slower or halt vehicle in front of it [7,12] in the same lane or during a cut-in (another vehicle comes in between the ACC and preceding vehicles, when the ACC vehicle is in vehicle- following mode) from another lane [20], or sudden braking applied by the preceding vehicle [16,18] or stop and go scenario [10,21,22]. During each TM, the ACC vehicle has to execute a high deceleration manoeuvre in order to avoid the crash with the preceding vehicle. The acceleration tracking capability of an ACC vehicle must be of high accuracy [8]. The acceleration tracking task is more challenging, because due to the deceleration limits an ACC vehicle is not capable of applying the required brake torque to evade a crash with any object in front of it, and this can cause the brake torque saturation [7,8]. The TM will be performed in the presence of acceleration, states and collision avoidance constraints when the brake and engine actuators have limited allowable forces that may saturate [7,8,18]. The development of the overall system model includes: vehicle modelling, controllers modelling, and their interaction.

## 2. TWO-VEHICLE SYSTEM MODEL

A two-vehicle system is considered which consists of a preceding vehicle and an ACC vehicle as shown in Fig. 1. The preceding vehicle travels independently, whereas, the ACC vehicle keeps a longitudinal distance from the preceding vehicle.

The longitudinal control of the ACC vehicle consists of two separate controllers as shown in Fig. 2. The ULC calculates the required acceleration commands for the LLC to maintain the required spacing behind the preceding vehicle. The LLC uses these desired acceleration commands to generate the required throttle and braking commands for the nonlinear ACC vehicle to follow the required acceleration commands calculated by the ULC [11].

The spacing policy between the two vehicles is based on the headway control policy. The headway time (h) can be defined as the time taken by the follower vehicle to reach the point where the preceding vehicle is at present speed [23]. The control loop diagram of the two vehicles is shown in Fig. 3. A first-order lag is considered in the ULC input command which correspond to the LLC's performance and comes from brake or engine actuation lags and sensor signal processing lags [8,16]. The first-order lag can be defined as [8,16]:

$$\tau \ddot{x}_2(t) + \dot{x}_2(t) = u(t) \tag{1}$$

where $x_1$ and $x_2$ are the absolute positions of the preceding and ACC vehicles. $u$ is defined as the control input commands determined by the ULC. $\tau$ is the time-lag equivalent to the lag in the LLC performance. Analytical and experimental studies show that has a value of 0.5s [8,24] and the same value is used in this study.

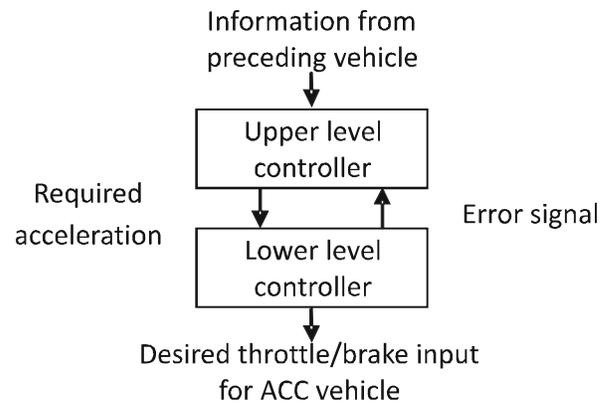

*FIG. 2. ACC VEHICLE LONGITUDINAL CONTROL SYSTEM [8]*

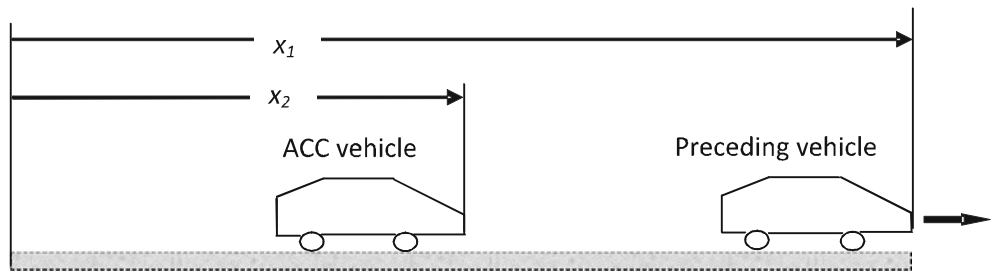

*FIG. 1. A TWO-VEHICLE SYSTEM*

Each vehicle's longitudinal motion is described in the continuous-time domain using a set of differential equations. Whereas, the MPC-based vehicle-following control laws for tracking the desired acceleration are calculated using discrete-time model.

## 2.1 Objectives

The paper presents the application of MPC technique to a nonlinear ACC vehicle model under TMs and examines the vehicle's sensitivity against a MPC parameter. The detailed study has been conducted in the Ph.D. project [25] and the simulation results have been compared with the suitable parameters [25] which ensure guaranteed response. The controller parameter selected is control input cost weighting coefficient. The paper outline is as follows. In Section 3 a nonlinear vehicle model to control the longitudinal dynamics of the vehicle is developed. Section 4 presents the MPC-based ULC's formulations which are used to control the longitudinal dynamics of the nonlinear ACC vehicle model. The derivation of the MPC based spacing-control law is presented which is used within the ULC formulation. Section 5 presents and discusses simulation results of the following ACC vehicle under TMs. The conclusions are provided in Section 6.

## 3. VEHICLE MODEL

A 3.8 litre spark-ignition engine model which consists of two states cylinders and a five-speed automatic transmission has been chosen, where the two states are the intake manifold pressure (p$man$) and the engine speed ($\omega_e$).

$$\dot{p}_{man} = \frac{RT_{man}}{V_{man}}(\dot{m}_{ai} - \dot{m}_{ao}) \tag{2}$$

$$I_e \dot{\omega}_e = T_i - T_f - T_a - T_p \tag{3}$$

where $T_{man}$ is the manifold temperature, $R$ is the universal gas constant of air, $V_{man}$ is the intake manifold volume,

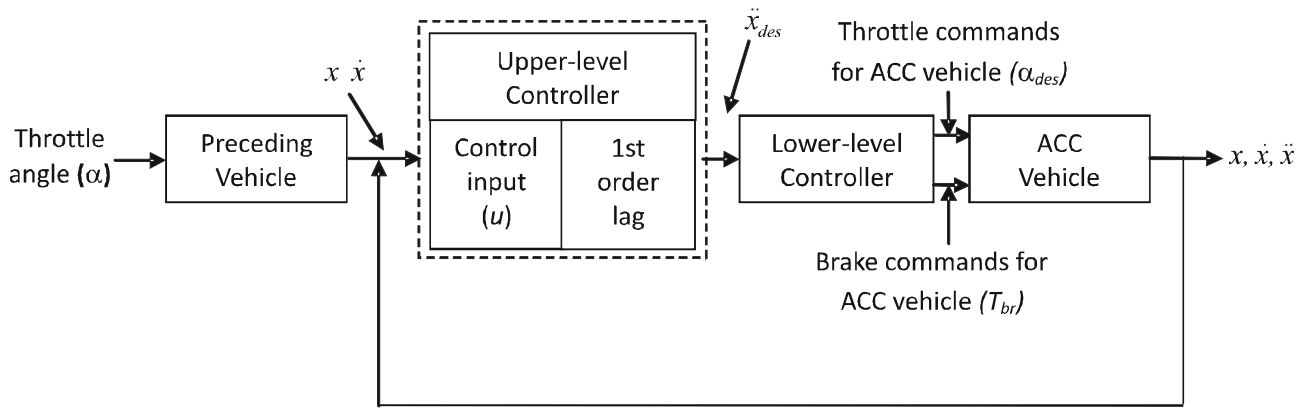

FIG. 3. CONTROL LOOP DIAGRAM FOR A 2-VEHICLE SYSTEM COMPRISES A PRECEDING AND AN ACC VEHICLE

$\dot{m}_{ai}$ and $\dot{m}_{ao}$ represent mass flow rate in and out of the intake manifold, $T_i$ is the engine combustion torque, $T_f$ is the engine friction torque [26], $T_a$ is the accessory torque, and $T_p$ is the pump torque representing the external load on the engine, and $I_e$ is the effective inertia of engine. The input to the engine model is the throttle angle.

Along the vehicle longitudinal axis a force balance results.

$$m\ddot{x} = F_{xf} - Faero - \frac{R_x}{r_{eff}} - mg\sin\theta \qquad (4)$$

where $m$ refers to the mass of the vehicle, $x$ is the vehicle displacement, $F_{xf}$ is the longitudinal tyre force at the front tire, $F_{aero}$ is aerodynamic drag force [26], $R_x$ is the rolling resistance torque [26], $r_{eff}$ is the effective tyre radius, and $\theta$ is gradient of the road. This nonlinear vehicle longitudinal dynamics model is used for both vehicles. The nonlinear vehicle model has been carefully redeveloped and assessed in [25,27] for the suitability of the two-vehicle system to control the longitudinal dynamics. The necessary parameters of the vehicle model are listed in Table 1, based on the information from [26].

## 4. MPC CONTROLLERS FORMULATION

The MPC-based ULC is presented in this paper and the details of the LLC algorithm can be found in [25]. This is one of the aims of this study. The ULC uses the range R and range rate $\dot{R}$ between both vehicles to determine the desired acceleration commands as shown in Fig. 4.

**TABLE 1. VEHICLE SYSTEM PARAMETERS**

| Engine Displacement | $V_d$ | 0.0038 m³ |
|---|---|---|
| Intake Manifold Volume | $V_{man}$ | 0.0027 m³ |
| Manifold Temperature | $T_{man}$ | 293 K |
| Engine Moment of Inertia | $I_e$ | 0.1454 kg.m² |
| Mass of the Vehicle | $m$ | 1644 kg |
| Accessory Torque | $T_a$ | 25 Nm |
| Effective Tyre Radius | $r_{eff}$ | 0.3 m |
| Wheel Moment of Inertia | $I_w$ | 2.8 kg.m² |

The main task by using the MPC method for the TM is to operate the system close to the constraint boundaries.

The main tasks for the MPC control method on the ACC system are to:

(i)     Track smoothly desired acceleration commands.

(ii)    Reach and maintain a SIVD in a comfortable manner and at the same time react quickly in the case of dangerous scenarios.

(iii)   Optimize the system performance within defined constrained operational boundaries.

There are some fundamental features of a MPC control algorithm which differentiate it from other control methods: its capability to develop explicitly a model to predict the process output at future time instants (horizon), the ability to design a reference path that the controller attempts to follow, calculation of a control sequence minimizing an objective function, and receding strategy; which means that at each instant the horizon moves forward to the future by applying the first control signal of the sequence calculated at each step [28].

### 4.1 Moving Horizon Window

The moving horizon window also referred as time-dependent window which can start from any arbitrary time $t_i$ to the prediction horizon $t_i+N_p$ for $i=1,...,N_p$. The $N_p$ prediction horizon ($N_p$) defines how far ahead in time the future output states are predicted and its length remains constant. However, $t_i$ which actually starts the optimization window, increases based on sampling instant [29].

### 4.2 Receding Horizon Control

The algorithm of the MPC controller is regarded as shown in Fig. 5. A discrete-time setting is assumed, and the current time is labelled as time step t. A set-point trajectory shown is the absolute target for the system to follow. It is unlikely that the system will follow exactly the set-point trajectory.

The reference path is therefore a newly defined path which starts from the current output at time t and defines an ideal path along which the plant (vehicle) should return to the set-point trajectory.

A MPC controller has an internal model which is used to predict the behaviour of the plant, starting at the current time t, over a future prediction horizon ($N_p$). Using the current output state information $y(t)$ of the system and with defined future control inputs $u(t+m/t)$ for m=0,..., $N_c$, the system's predicted outputs $y(t+m/t)$ for m=1,..., $N_p$ are obtained up to a limited prediction horizon ($N_p$) [28-30].

The set of future control inputs $u(t+m/t)$ for m=0,..., $N_c$ is determined up to the control horizon ($N_c$), Fig. 5, by optimizing the suitable measure (determined criterion) to maintain the process close to the reference path [31]. This criterion usually represented as a quadratic function of the errors (between the predicted output signal and the predicted reference trajectory), also taking into account the control effort input. Changes in the control input are weighted and accumulated in the quadratic function. During this process an online computation is used to find out the state-trajectories which are actually based on the current state and then a cost minimizing control strategy is determined until time $t+N_c$.

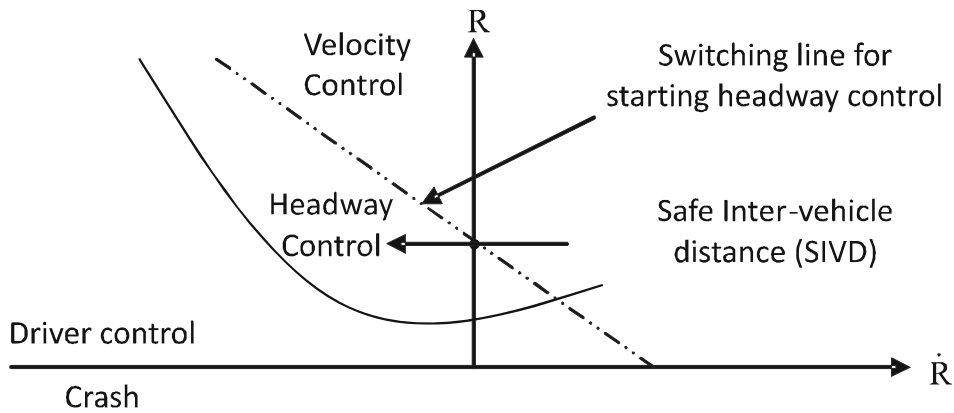

*FIG. 4. RANGE VS. RANGE-RATE DIAGRAM [8]*

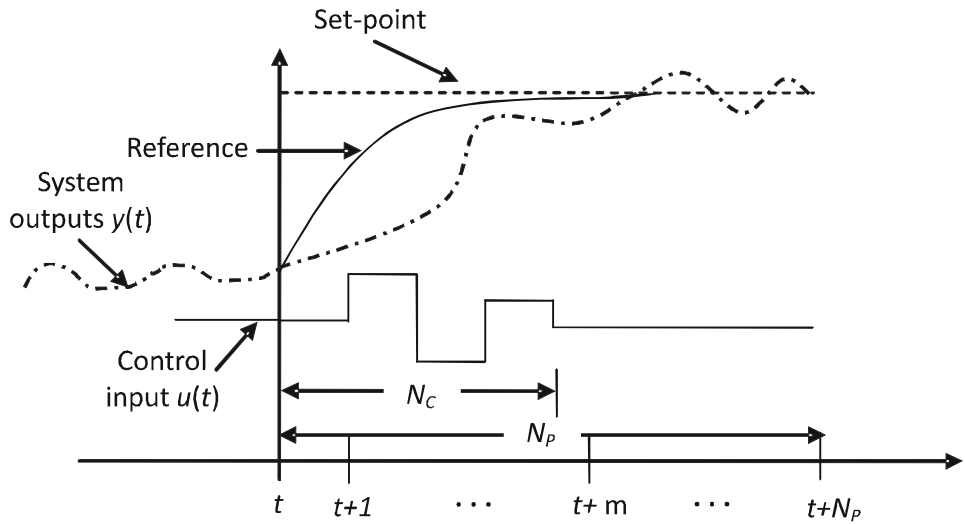

*FIG. 5. MPC STRATEGY: BASIC IDEA [30]*

Once the future control inputs are determined then only the first element of the set of future control inputs is applied as the input signal to the plant. During this process the prediction horizon length remains constant as before, but glide by one time interval at each step, this entire phenomenon is called a receding horizon strategy [29,30].

## 4.3    Cost Function and Control Objective

At each time t, the state information is sampled in order to predict the future control strategy. Once the sampling process is completed this information is then compared with the desired value (reference path), this comparison then generates an error function which is based on the difference of these two values. This error function is formulated as a cost function, '*J*', which consists of elements relating to the system's output accuracy and control effort input. The cost function also incorporates the weighting which penalizes the control input u(t) for the required performance of closed-loop. The control objective is to minimize *J* inside the optimization window and by doing so the optimized control action is determined [29].

## 4.4    Formulation of Prediction Model

For the purpose of illustration of the MPC control algorithm a linearized, continuous-time, SISO (Single Input and Single Output) system is considered and is described by:

$$\dot{x} = Ax + Bu \qquad (5)$$

$$y = Cx + Du \qquad (6)$$

where x represents the state variable, u denotes the control input, *y* refers to the system output, and A,B,C,D are the state-space matrices. The system matrix D is assumed zero because the control input u has no influence on the output *y* due to receding horizon control principle [29].

In the MPC literature, controlled system is usually modelled by a discrete time state space model [16,32]. Therefore, the continuous time state space model, Equations (5-6), is altered into a discrete time state space model as:

$$x(k+1) = \mathbf{A}x(k) + \mathbf{B}u(k) \qquad (7)$$

$$y(k) = \mathbf{C}x(k) \qquad (8)$$

where k represents the kth sampling point. The prediction is performed within an optimization window $N_p$ which is the number of samples and each sample is denoted by the time $k_t$, $k_t > 0$. At each time instant $k_t$, the state-vector x($k_i$) is measured which provides the current plant information. Having the current plant state x($k_i$), the upcoming states are then envisaged for $N_p$ instants and the future state variables can be defined as:

$$x(k_i+1|k_i), x(k_i+2|k_i),...,x(k_i+m|k_i),...,x(k_i+N_P|k_i) \qquad (9)$$

where x($k_i$+m/$k_i$) is the envisaged state-variable at $k_i$+m with the given recent state x($k_i$). Similarly, using the recent system state x($k_i$), the set of future control input, which minimizes the cost function *J*, are denoted by:

$$\Delta u(k_i), \ \Delta u(k_i+1),......,\Delta u(k_i+N_c-1) \qquad (10)$$

where $\Delta u(k)$ is the control increment (augmented model). $N_c$ is called the length of control-horizon [30]. The length of $N_c$ should be less than or equal to the length of $N_p$.

The future state variable in Equation (9) can be calculated sequentially using the current state vector and the set of future control parameters.

$$x(k_i+1|k_i) = \mathbf{A}x(k_i) + \mathbf{B}\Delta u(k_i)$$

$$x(k_i+2|k_i) = \mathbf{A}x(k_i+1) +$$

$$\Delta u(k_i+1) = \mathbf{A}^2 x(k_i) + \mathbf{AB}\Delta u(k_i) + \mathbf{B}\Delta u(k_i+1)$$

$$\vdots$$

$$x(k_i+N_P|k_i) = \mathbf{A}^{N_P} x(k_i) +$$

$$\mathbf{A}^{N_P-1}\mathbf{B}\Delta u(k_i) + \mathbf{A}^{N_P-2}\mathbf{B}\Delta u(k_i+1)$$

$$+,....,+\mathbf{A}^{N_P-N_C}\mathbf{B}\Delta u(k_i+N_C-1)$$

$$(11)$$

Similarly, using Equation (11) the foreseen output variables can be determined as:

$$y(k_i + 1 \mid k_i) = \mathbf{CA}x(k_i) + \mathbf{CB}\Delta u(k_i)$$

$$y(k_i + 2 \mid k_i) = \mathbf{CA}x(k_i + 1) + \mathbf{B}\Delta u(k_i + 1) = \mathbf{CA}^2 x(k_i)$$

$$+ \mathbf{CAB}\Delta u(k_i) + \mathbf{CB}\Delta u(k_i + 1)$$

$$y(k_i + 3 \mid k_i) = \mathbf{CA}^3 x(k_i) + \mathbf{CA}^2 \mathbf{B}\Delta u(k_i)$$

$$+ \mathbf{CAB}\Delta u(k_i + 1) + \mathbf{CB}\Delta u(k_i + 2)$$

$$\vdots \tag{12}$$

$$y(k_i + N_P \mid k_i) = \mathbf{CA}^{N_P} x(k_i) + \mathbf{CA}^{N_P - 1}\mathbf{B}\Delta u(k_i)$$

$$+ \mathbf{CA}^{N_P - 2}\mathbf{B}\Delta u(k_i + 1)$$

$$+ \ldots + \mathbf{CA}^{N_P - N_C}\mathbf{B}\Delta u(k_i + N_C - 1)$$

The above equations can be written in the vector form as:

$$\mathbf{Y} = \begin{bmatrix} y(k_i+1 \mid k_i) & y(k_i+2 \mid k_i) & y(k_i+3 \mid k_i) & \ldots & y(k_i+N_P \mid k_i) \end{bmatrix}^T \tag{13}$$

$$\Delta\mathbf{U} = \begin{bmatrix} \Delta u(k_i) & \Delta u(k_i+1) & \Delta u(k_i+2) & \ldots & \Delta u(k_i+N_C-1) \end{bmatrix}^T \tag{14}$$

where the length of $\mathbf{Y}$ is equal to $N_p$ and the length of $\Delta\mathbf{U}$ is equal to $N_C$. Equations (13-14) can be re-written into a state space expression, calculating all system outputs using the initial states $x(k_i)$ and vector of predicted control inputs $\Delta\mathbf{U}$ as:

$$\mathbf{Y} = \mathbf{F}x(k_i) + \mathbf{\Phi}\Delta\mathbf{U} \tag{15}$$

where

$$\mathbf{F} = \begin{bmatrix} \mathbf{CA} \\ \mathbf{CA}^2 \\ \mathbf{CA}^3 \\ \vdots \\ \mathbf{CA}^{N_p} \end{bmatrix} \tag{16}$$

and

$$\mathbf{\Phi} = \begin{bmatrix} \mathbf{CB} & 0 & 0 & \ldots & 0 \\ \mathbf{CAB} & \mathbf{CB} & 0 & \ldots & 0 \\ \mathbf{CA}^2\mathbf{B} & \mathbf{CAB} & \mathbf{CB} & \ldots & 0 \\ \vdots & & & & \\ \mathbf{CA}^{N_P-1}\mathbf{B} & \mathbf{CA}^{N_P-2}\mathbf{B} & \mathbf{CA}^{N_P-3}\mathbf{B} & \ldots & \mathbf{CA}^{N_P-N_C}\mathbf{B} \end{bmatrix} \tag{17}$$

For detailed understanding of the augmented model, the discrete time state space model (Equation (7)) and its transformation into the state-space model (Equation (15)), the reader is referred to Maciejowski [30] and Wang [29].

## 4.5 Control Input Optimization

The cost function $J$, that describes the control objective, can be defined as:

$$J = (\mathbf{R}_S - \mathbf{Y})^T (\mathbf{R}_S - \mathbf{Y}) + \Delta\mathbf{U}^T \overline{\mathbf{R}} \Delta\mathbf{U} \tag{18}$$

The cost function $J$ consists of two separate terms. The first terms is meant to minimize the error between desired output and the predicted output while the second term deals with the size of $\Delta\mathbf{U}$ when the cost function J is made as small as possible. $\overline{\mathbf{R}} = \mathbf{R}\mathbf{I}_{N_C \times N_C} (\mathbf{R} \geq 0)$ where $\mathbf{R}$ is employed as a fine-tuning operator for the needed closed-loop performance [29] which penalizes the control input vector ($\Delta\mathbf{U}$). $\mathbf{R}_s$ is the vector that contains the desired state information and can be defined as:

$$\mathbf{R}_S^T = \overbrace{\begin{bmatrix} 1 & 1 & \ldots & 1 \end{bmatrix}}^{N_P} r(k_i) \tag{19}$$

where $r(k_i)$ is the given set-point signal at time instant $k_i$ [28].

The next step is to find $\Delta\mathbf{U}$ which can be obtained by substituting $\mathbf{Y}$ in Equation (18) and re-arranging as:

$$J = (\mathbf{R}_S - \mathbf{F}x(k_i))^T (\mathbf{R}_S - \mathbf{F}x(k_i)) - 2\Delta\mathbf{U}^T \mathbf{\Phi}^T (\mathbf{R}_S - \mathbf{F}x(k_i)) + \Delta\mathbf{U}^T (\mathbf{\Phi}^T \mathbf{\Phi} + \overline{\mathbf{R}})\Delta\mathbf{U} \tag{20}$$

Taking the 1st derivative of $J$

$$\frac{\partial J}{\partial \Delta\mathbf{U}} = -2\mathbf{\Phi}^T (\mathbf{R}_S - \mathbf{F}x(k_i)) + 2(\mathbf{\Phi}^T \mathbf{\Phi} + \overline{\mathbf{R}})\Delta\mathbf{U} \tag{21}$$

And the required condition for the minimized $J$ can be expressed as:

$$\frac{\partial J}{\partial \Delta\mathbf{U}} = 0 \tag{22}$$

From which $\Delta\mathbf{U}$ can be optimized as:

$$\Delta\mathbf{U} = (\mathbf{\Phi}^T\mathbf{\Phi} + \overline{\mathbf{R}})^{-1}\mathbf{\Phi}^T(\mathbf{R}_s - \mathbf{F}\mathbf{x}(k_i)) \qquad (23)$$

Once the control parameter vector is calculated then only the first element is applied to the controlled system.

## 4.6 MPC Prediction Model for the Two-Vehicle System

In this section the MPC control algorithm is applied to the two-vehicle system which consists of a preceding vehicle and a following ACC vehicle. Both vehicles are based on the nonlinear vehicle model (Section 3) and the longitudinal states of both vehicles are obtained by taking the force balance along the vehicle longitudinal axis [25] (for both vehicles separately). The position of the preceding vehicle is denoted by $x_1$ and the position of the ACC vehicle is denoted by $x_2$ (Fig. 1).

The continuous-time model in Equation (1) can be re-written in a discrete-time state-space model as [7]:

$$\begin{pmatrix} x_2(t+T) \\ \dot{x}_2(t+T) \\ \ddot{x}_2(t+T) \end{pmatrix} = \begin{pmatrix} 1 & T & 0 \\ 0 & 1 & T \\ 0 & 0 & 1-\dfrac{T}{\tau} \end{pmatrix} \begin{pmatrix} x_2(t) \\ \dot{x}_2(t) \\ \ddot{x}_2(t) \end{pmatrix} + \begin{pmatrix} 0 \\ 0 \\ \dfrac{T}{\tau} \end{pmatrix} u(t) \qquad (24)$$

where $T$ is the discrete sampling time of the ACC system and assumed as 0.1s.

### 4.6.1 Coordinate Frame for Transitional Manoeuvres

It is of paramount importance to develop and understand the mathematical relation between the state variables of both vehicles. The desired SIVD between the two vehicles varies linearly with the preceding vehicle's speed such that the headway time ($h$) between the two vehicles remains constant (SIVD=$hv_{preceding}$).

A coordinate frame [7] as shown in Fig. 6, moves with a speed equal to the preceding car velocity. This frame is used to determine the ACC vehicle motion relative to the preceding car. The origin of this frame is situated at the desired SIVD and the goal of TM is to steer the ACC vehicle to the origin of this frame in order to set up the zero range-rate with the preceding vehicle, where, R is the range (relative distance) between the two vehicles.

Using this coordinate frame (Fig. 6) for TM, the discrete-time state-space model of the error vector between the two vehicles can be defined as:

$$\mathbf{e}_{k+1} = \mathbf{A}\mathbf{e}_k + \mathbf{B}u_k \qquad (25)$$

$$y_k = \mathbf{C}\mathbf{e}_k \qquad (26)$$

where,

$$\mathbf{e}_k = \begin{pmatrix} err \\ \dot{err}_k \\ \ddot{err}_k \end{pmatrix} = \begin{pmatrix} -(R-SIVD) \\ \dot{R} \\ \ddot{x}_k \end{pmatrix} \qquad (27)$$

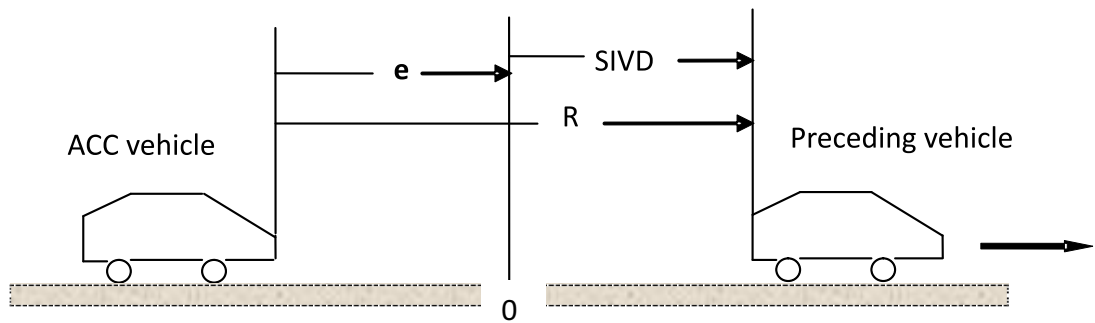

*FIG. 6. COORDINATE FRAME FOR TRANSITIONAL MANOEUVRE.*

where, $err_k$ is spacing error, $\dot{err}_k$ is range-rate (relative velocity between the two vehicles), and $\ddot{err}_k$ is the absolute acceleration of the ACC vehicle. Each element of the error vector ($\mathbf{e}_k$) is the quantity which is measured by the ACC system and the control objective is to steer these quantities to zero [7]. $u_k$ is the control input, and $y_k$ is the system output at time step $k$. The system matrices $\mathbf{A}$ and $\mathbf{B}$ can be obtained from the comparison of Equations (24-25).

$$\mathbf{A} = \begin{pmatrix} 1 & T & 0 \\ 0 & 1 & T \\ 0 & 0 & 1-\dfrac{T}{\tau} \end{pmatrix}, \qquad \mathbf{B} = \begin{pmatrix} 0 \\ 0 \\ \dfrac{T}{\tau} \end{pmatrix} \qquad (28)$$

And the system matrix C is defined as [7]:

$$\mathbf{C} = \begin{pmatrix} 1 & 0 & 0 \\ 0 & -1 & 0 \end{pmatrix} \qquad (29)$$

Using the MPC control approach, Section 4.4, the future error can be defined as:

$$\mathbf{e}(k_i+1\,|\,k_i), \mathbf{e}(k_i+2\,|\,k_i),\dots,\mathbf{e}(k_i+\mathrm{m}\,|\,k_i),\dots,\mathbf{e}(k_i+N_P\,|\,k_i) \qquad (30)$$

Where $\mathbf{e}(k_i+m/k_i)$ is the predicted error variable at $k_i+$m with the given current error information $\mathbf{e}(k_i)$. The set of future control inputs are denoted by:

$$\Delta u(k_i),\ \Delta u(k_i+1),\dots\dots\Delta u(k_i+N_C\text{-}1) \qquad (31)$$

where $\Delta u(k)=u(k)-u(k\text{-}1)$ is the control increment. The MPC controller forms a cost function (Equation (18)) which consists of these errors ($\mathbf{e}_k$) and the control input which determines the best set of future control inputs to balance the output error minimisation against control input effort. Repeating the derivation steps from Equation (11) to Equation (23) one can find the optimal solution for the control input which is the function of current error information $\mathbf{e}(k_i)$.

$$\Delta\mathbf{U} = (\mathbf{\Phi}^T\mathbf{\Phi} + \bar{\mathbf{R}})^{-1}\mathbf{\Phi}^T\,(\mathbf{R}_s - \mathbf{Fe}_{k_i}) \qquad (32)$$

For this study, $r(k_i)=0$, because the control objectives are to steer the error vector ($\mathbf{e}_k$) to 0, i.e. the spacing error ($err_k$) should steer to zero so the desired SIVD could be achieved and range-rate $\left(\dot{err}_k\right)$ should converge to zero so the ACC vehicle follow the ACC vehicle with the same velocity. The absolute acceleration $\left(\ddot{err}_k\right)$ should steer to zero so the ACC vehicle moves with the constant speed. For the predictive model developed $\mathbf{R}=1$. $\mathbf{F}$ and $\mathbf{\Phi}$ matrices are defined in Equation (16-17) respectively.

At each time step $k$ the MPC algorithm determines a sequence of control inputs ($\Delta\mathbf{U}_0\dots\Delta\mathbf{U}_{NC\text{-}1}$) to minimize the cost function $J$ (Equation (18)). The parameters which have been used in the MPC controller formulation are shown in Table 2.

During the control algorithm formulation, the operational constraints are incorporated in the MPC controller formulation. The constraints incorporated are control input constraint, Equation (33). State constraint which means the ACC vehicle cannot have a negative velocity. The accident prevention has also been put together as state constraint, Equation (35). And the terminal constraint which refers the ACC vehicle should establish a SIVD with the zero range-rate.

**TABLE 2. CONTROLLER PARAMETERS**

| Discrete Time Sample | $T$ | 0.1s |
|---|---|---|
| Time Lag | $\tau$ | 0.5s |
| Tuning Operator | $\boldsymbol{R}$ | 1 |
| Set Point | $r$ | 0 |
| Headway Time | $h$ | 1s |
| Prediction Horizon | $N_p$ | 230 Samples |
| Control Horizon | $N_c$ | 3 Samples |
| Upper Acceleration Limit | $u_{max}$ | $0.25g$ |
| Lower Acceleration Limit | $u_{min}$ | $-0.5g$ |

The control input constraint included in the MPC control formulation is:

$$u_{\min} \leq u_k \leq u_{\max} \tag{33}$$

The dimension of $\Delta\mathbf{U}$ is $N_C$ and $N_C$ is 3 samples, therefore, the constraints are fully imposed on all the components in $\Delta\mathbf{U}$ and can be translated to the six linear inequalities as:

$$\begin{bmatrix} 1 & 0 & 0 \\ 0 & 1 & 0 \\ 0 & 0 & 1 \\ -1 & 0 & 0 \\ 0 & -1 & 0 \\ 0 & 0 & -1 \end{bmatrix} \begin{bmatrix} \Delta u(k_i) \\ \Delta u(k_i+1) \\ \Delta u(k_i+2) \end{bmatrix} \leq \begin{bmatrix} u_{\max} - u(k_i-1) \\ u_{\max} - u(k_i-1) \\ u_{\max} - u(k_i-1) \\ u_{\min} + u(k_i-1) \\ u_{\min} + u(k_i-1) \\ u_{\min} + u(k_i-1) \end{bmatrix} \tag{34}$$

And the state and collision avoidance constraints incorporated in the MPC control formulation are:

$$y_k = \begin{pmatrix} err_k \\ -\dot{err}_k \end{pmatrix} \leq \begin{pmatrix} \text{SIVD} \\ v_{preceding} \end{pmatrix} \tag{35}$$

$$\text{SIVD} = h v_{\text{preceding}} \tag{36}$$

where, $h$ is the headway time.

## 5.    RESULTS AND DISCUSSION

This section presents the simulation analyses of the nonlinear ACC vehicle for different values of control input cost weighting (**R**) under the critical TM. The control objectives for the complex ACC vehicle are; "to perform the critical TMs (for deceleration limit of -0.5$g$) in order to establish and maintain the SIVD with the zero range-rate behind a newly detected slower or halt preceding vehicle". The TMs will be performed in the presence of acceleration constraint (Equation (33)), states and accident prevention constraints (Equation (35)).

## 5.1    Sensitivity Analysis for Different Control Input Cost Weighting Coefficient (R)

In this scenario an ACC vehicle travelling at a speed of 30 m/s detects an accelerating preceding vehicle. The nonlinear ACC vehicle model has to decelerate from 30 m/s to the velocity of the preceding vehicle. The initial range between the two vehicles is 60m. The initial velocity of the preceding vehicle is 10 m/s. The corresponding initial engine speed, throttle input, and gear ratio for the preceding vehicle are 1967 rpm, 50 degrees, 2$^{nd}$ gear, and for ACC vehicle are 4040 rpm, 70 degrees, and 5$^{th}$ gear, respectively.

The reaction of the ACC vehicle to the MPC controller is shown in Fig. 7 where both vehicles are based on nonlinear vehicle models. Vehicle positions (Fig. 7(a)), velocities (Fig. 7(b)), accelerations (Fig. 7(c)), engine speeds (Fig. 7(d)), range (Fig. 7(i)), for the two vehicles have been plotted. The accelerating preceding vehicle starting from 10 m/s velocity and 2$^{nd}$ gear at 50 degree throttle input reaches up to 31.3 m/s (112.68 km/h) as shown in Fig. 7(b). At this speed the operating gear ratio of the preceding vehicle is 3$^{rd}$ gear and the throttle input is constant all the time.

The main purpose of this analysis is to analyse how the ACC vehicle will respond for different values of **R** and how its values affects the engine and transmission dynamics. Two different values of **R** other than 1 are used for the sensitivity analysis purpose; **R**=0.1 and **R**=20. The lower value of **R** represents a looser control and the higher value of **R** represents a tighter control. Results for different values of **R** have been shown in Fig. 7; in the case of lower value the ACC vehicle response shows a higher disturbance in the ACC vehicle's response. The ACC vehicle is unable to establish a SIVD (Fig. 7(e)) and cannot setup a zero range-rate (Fig. 7(b-c)) demonstrates that the ACC vehicle maintains the deceleration limit (-0.5$g$) but it cannot avoid the collision with the preceding vehicle. The engine and transmission response (Fig. 7(d)) are not catching up with the preceding vehicle response at all. Fig. 7(e) shows that during the transitional operation the ACC vehicle has collided with the preceding vehicle and during the steady-state operation it is travelling far behind the preceding vehicle. This analysis with lower value of **R** (**R**=0.1) shows the computed control input signal is not enough to control the dynamic behaviour of the ACC vehicle.

In the case of higher **R**=20 value the ACC vehicle response is quite satisfactory. It has been precisely observed that during the transitional operation and steady-state operation the response of the ACC vehicle has not been delayed but has been prolonged. This is because of the higher influence of the cost weighting on the optimal control input. The ACC vehicle can successfully perform the TM and can achieve the desired control objectives. The analysis carried out in this section shows that a lower value of **R**(**R**<1) for input is not suitable for the ACC vehicle at all when using the MPC control algorithm, however, a higher value up to 20 can be used for the ACC vehicle in order to improve the performance of the ACC vehicle.

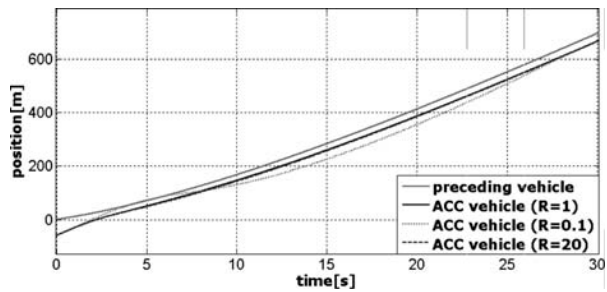

*(a) POSITIONS OF BOTH VEHICLES*

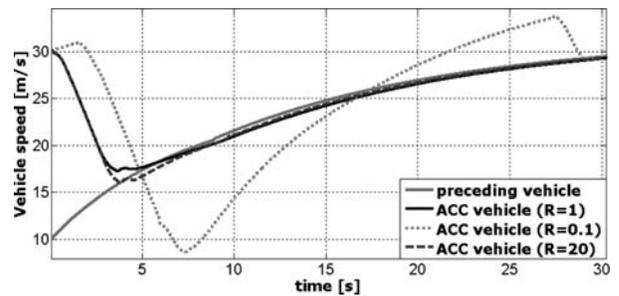

*(b) VELOCITIES OF BOTH VEHICLES*

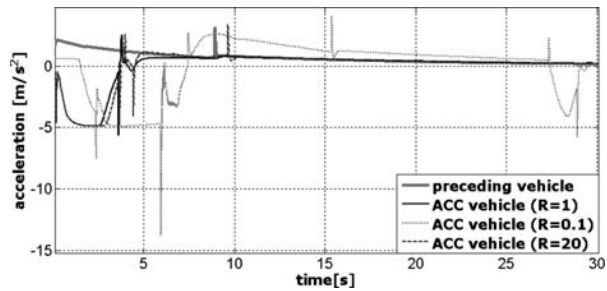

*(c) ACCELERATIONS OF BOTH VEHICLES*

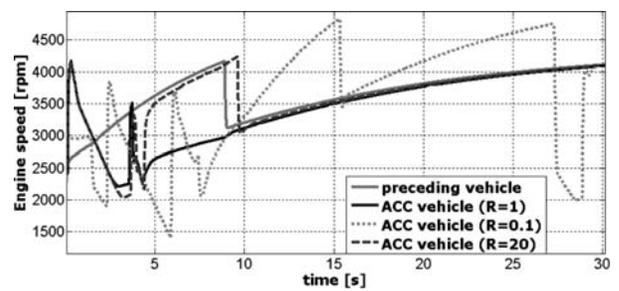

*(d) ENGINE SPEED*

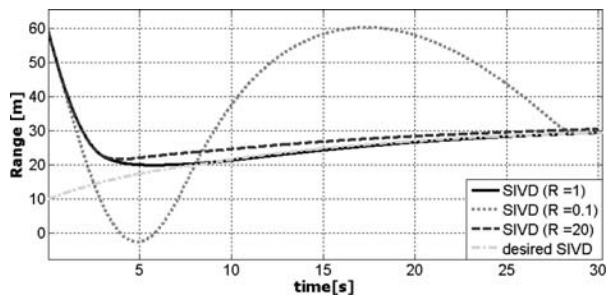

*(e) RANGE*

*FIGURE 7. RESPONSE OF ACC VEHICLE FOR DIFFERENT VALUES OF R*

# 6.    CONCLUSIONS

An application of mathematical control techniques to the longitudinal dynamics of a nonlinear vehicle equipped with an ACC system is presented. The vehicle model captures the vehicle dynamics even during the transmission gear shifting. A comparison of the above results with previous results shows that the nonlinear vehicle is highly sensitive to values of **R**<1. The resulting control action taken by the controller is no longer sufficient and seriously influences the vehicle's inherent dynamics. The driving experience highlights delays in achieving the desired levels. However, a higher value of **R**(**R**>1) is found promising and guarantees the required performance. It has been observed that ACC vehicle effectively performs the required TM, avoids the collision with the preceding vehicle, and set up the desired SIVD and establishes the zero range-rate. It should be noted that the ACC vehicle is obeying all the applied constraints during this TM, i.e. control input, states, and collision avoidance while the constraints are applied in the ULC formulation only.

## ACKNOWLEDGMENTS

The authors would like to express their thanks to Mehran University of Engineering & Technology, Jamshoro, Pakistan, for giving yhem an opportunity to pursue  this study. The reserach work/study was supported by Higher Education Commission, Government of Pakistan, under Faculty Development Scheme.